\documentclass[aps,prl,twocolumn,longbibliography,superscriptaddress]{revtex4-1}
\usepackage{xcolor}
\usepackage[normalem]{ulem}
\usepackage{graphicx}
\usepackage{dcolumn}
\usepackage{amsmath}
\usepackage{amssymb}
\usepackage{bm}

\begin{document}

\title{Irreversibility across a nonreciprocal ${\cal PT}$-symmetry-breaking phase transition}

	\author{Henry Alston}
	\affiliation{Department of Mathematics, Imperial College London, South Kensington, London SW7 2AZ, United Kingdom}

	\author{Luca Cocconi}
	\affiliation{Department of Mathematics, Imperial College London, South Kensington, London SW7 2AZ, United Kingdom}
	\affiliation{The Francis Crick Institute, London NW1 1AT, United Kingdom}
	\affiliation{Max Planck Institute for Dynamics and Self-Organization (MPIDS), 37077 G\"{o}ttingen, Germany}
			
	\author{Thibault Bertrand}%
	\email{t.bertrand@imperial.ac.uk}
	\affiliation{Department of Mathematics, Imperial College London, South Kensington, London SW7 2AZ, United Kingdom}

\date{\today}


\begin{abstract}
Nonreciprocal interactions are commonplace in continuum-level descriptions of both biological and synthetic active matter, yet studies addressing their implications for time-reversibility have so far been limited to microscopic models. Here, we derive a general expression for the average rate of informational entropy production in the most generic mixture of conserved phase fields with nonreciprocal couplings and additive conservative noise. For the particular case of a binary system with Cahn-Hilliard dynamics augmented by nonreciprocal cross-diffusion terms, we observe a non-trivial scaling of the entropy production rate across a parity-time symmetry breaking phase transition. We derive a closed-form analytic expression in the weak-noise regime for the entropy production rate due to the emergence of a macroscopic dynamic phase, showing it can be written in terms of the global polar order parameter, a measure of parity-time symmetry breaking.
\end{abstract}

\maketitle

Though the action-reaction principle implies reciprocity at the level of microscopic forces, effective reciprocity-breaking interactions commonly arise at the mesoscopic scale. In living matter, one may even argue that reciprocity-breaking interactions are the rule rather than the exception, as exemplified by the classical predator-prey and promoter-inhibitor models \cite{Mobilia2007, Biktashev2009, Hong2011, Lavergne2019}. A lack of reciprocity naturally comes up in systems whose dynamics is dependent on information propagation as in crowds of social animals \cite{Helbing2000,Vicsek2012,Bain2019,Paoluzzi2020}. Furthermore, nonreciprocal interactions generically emerge from microscopic interactions mediated by a nonequilibrium medium \cite{Ivlev2015, Durve2018, Hayashi2006,Paoluzzi2018}, leading to fundamentally nonequilibrium physics \cite{Fruchart2021, Kryuchkov2018, Bowick2022, Shankar2022}. 

The breaking of reciprocal-symmetry at the level of effective physical interactions generically leads to directed motion as seen in diffusiophoretic colloidal mixtures \cite{Saha2019, Soto2014, AgudoCanalejo2019} and binary systems of active and passive particles \cite{You2020,Mandal2022}. Self-propelling mesoscopic agents have been shown to naturally emerge from an imbalance of attraction-repulsion interactions between microscopic agents \cite{Meredith2020,Cocconi2023}. In biology, the \textit{chase-and-run} behavior displayed by neural crest cells and placodal cells provides a generic mechanism of coordinated cell migration which is key to many fundamental morphogenetic and physiological processes \cite{Theveneau2013}. Such nonreciprocal phase transitions have recently been formalized in \cite{Fruchart2021}. In particular, the emergence of dynamical phases in systems with nonreciprocal interactions has been associated with the breaking of parity-time (${\cal PT}$) symmetry \cite{You2020,Fruchart2021}, providing one more example to a wider class of ${\cal PT}$-symmetry breaking transitions \cite{ElGanainy2018}, which includes optical systems \cite{Lin2011}, directional interface growth \cite{Coullet1989,Cummins1993,Pan1994} and more recently polar swarm models \cite{Borthne2020,Fruchart2021}.

While the coarse-graining of a microscopic model is generally arduous, the top-down approaches based on conservation and symmetry principles used to model dynamic critical phenomena at equilibrium can be extended to active systems \cite{Cahn1958, Hohenberg1977, Bray2002}, leading to so-called active field theories \cite{Cates2022}. Scalar active field theories, such as \textit{Active Model B} \cite{Wittkowski2014} and \textit{B+} \cite{Tjhung2018}, have for instance been used to describe nonequilibrium liquid-gas phase separation phenomena \cite{Fausti2021, Alston2022a}. Owing to their simplicity and generality, these active field theories present an attractive starting point for analyzing the nonequilibrium thermodynamic properties of living systems. However, at the level of these continuum descriptions, the dynamics are formulated in terms of macroscopic order parameters (such as the density) and no direct connection can be made between the degree to which global detailed balance is broken and the rate of energy dissipation at the microscopic scale as the notion of particle entity was entirely lost \cite{Cocconi2021,Bothe2022}. Despite this, the extent to which time-reversal symmetry (TRS) is broken (often referred to as the rate of \textit{informational} entropy production) still provides a useful metric for measuring the \textit{distance from equilibrium} in nonequilibrium systems at the level of the macroscopic dynamics \cite{Fodor2016, Kullback1951, Seifert2012, Cocconi2020, Cocconi2021, Nardini2017, Li2021, Fodor2022}. Of particular interest are the effects of phase transitions in active systems, where discontinuities in the scaling of entropy production have recently been observed \cite{Yu2022,Ferretti2022}.

In this Letter, we elucidate the impact of nonreciprocal interactions and its associated ${\cal PT}$-symmetry breaking transition to a dynamic state on the time-reversibility of a nonreciprocal active field theory. First, we derive a general result for the average rate of informational entropy production in a mixture of scalar fields with mass-conserving dynamics by evaluating the Kullback-Leibler divergence per unit time of the ensemble of forward paths and their time-reversed counterparts \cite{Kullback1951,Gaspard2004}. This result is then applied to a field theory including nonreciprocal interactions introduced through linear asymmetric couplings \cite{Saha2020}. In the particular case of a binary system, for which the nonreciprocity is controlled by a single parameter, we show that the entropy production rate exhibits non-trivial scaling driven by a phase transition from an {isotropic} static to a {polar} dynamic phase. We further identify analytically the contribution to the entropy production coming from the emergence of macroscopic dynamics and show that it scales quadratically with the speed of the dynamic phase, strikingly mirroring the relation between self-propulsion speed and TRS-breaking observed in microscopic active systems \cite{Seifert2012, Cocconi2020, GarciaMillan2021}.

\textit{Informational entropy production in scalar active mixtures. ---} We consider a system of $N$ interacting, conserved active fields $\{\phi_i(\mathbf{r},t)\}_{i\in[1,N]}$, whose governing equations are of the form 
\begin{equation}\label{eq:eom}
	\dot{\phi}_i(\textbf{r}, t) = \nabla^2 \mu_i(\textbf{r}, t) + \nabla \cdot \mathbf{\Lambda}_i(\textbf{r}, t),
\end{equation}
where $\mu_i(\textbf{r}, t)$ is a chemical potential which can include passive as well as active contributions and $\mathbf{\Lambda}_i$ is a noise term capturing thermal fluctuations in the system.  For the sake of tractability, the noise is taken to be additive as is commonly done in field theories of active phase separation \cite{Tjhung2018, Wittkowski2014, Cates2022, Saha2019, Li2020, 
Li2021} for which the order parameter describes the fluctuations around a homogeneous state and hence the leading-order contribution to the noise correlator is homogeneous in space, at least in the disordered phase. 
In practice, the conserved noise terms appearing in Eq.\,(\ref{eq:eom}) require careful regularization since their power spectrum is unbound in the ultraviolet, which may lead to divergences. While this is often done by regularizing the noise correlator (without affecting the conservative nature of the noise), we show in the present work that divergences in the entropy production rate originate from the infinite dimensionality of the continuum field and can only be cured by imposing finite dimensionality \cite{Pruessner2022}. To do so, we employ a suitable spatial discretization scheme when analyzing the dynamics below, effectively imposing a UV cutoff (see \footnote{See Supplemental Material at []} for an extended discussion). Note that we keep notation pertaining to continuous space here for readability \cite{Tauber2014}.

For our system, the extent of time-reversal symmetry breaking is quantified by the Kullback-Leibler divergence per unit time \cite{Kullback1951, Nardini2017}:
\begin{equation}\label{eq:epr}
	\dot{\mathcal{S}} = \lim_{\tau\rightarrow\infty} \frac{1}{\tau} \bigg\langle \log \frac{\mathbb{P}_F[\{\phi_{i=\{1, \dots, N\}}\}_{t=0}^{t=\tau}]}{\mathbb{P}_B[\{\phi_{i=\{1, \dots, N\}}\}_{t=0}^{t=\tau}]}\bigg\rangle, 
\end{equation}
where the $\mathbb{P}_F[\cdot]$ and $\mathbb{P}_B[\cdot]$ denote the path-probability for the forward and backward paths, respectively, for the combined dynamics of the $N$ fields. The average of the $\log$-ratio is taken over realizations of the noise terms $\mathbf{\Lambda}_{i=\{1, \dots, N\}}$. In a thermodynamically consistent microscopic theory, $\dot{\mathcal{S}}$ would correspond to the total rate at which entropy is produced. However, here it shall be understood only in the informational sense, i.e. as a measure of TRS breaking in the dynamics, as argued above.

Employing the usual approach for the treatment of stochastic field theories, we know that the two path-probabilities can each be written in terms of a dynamical action which takes the form of an Onsager-Machlup functional \cite{Nardini2017,Seifert2012,Tauber2014}. Taking the log-ratio of these two path-probabilities, we see that Eq.\,(\ref{eq:epr}) is the difference of two Onsager-Machlup functionals. Suppose that the noise terms $\mathbf{\Lambda}_i$ in Eq.\,(\ref{eq:eom}) are independent, with a diagonal correlation matrix $\Theta_{ij}=\langle \Lambda_i(\textbf{r}, t)\Lambda_j(\textbf{r}', t')\rangle = 2D\delta_{ij}\delta(\mathbf{r}'-\mathbf{r})\delta(t'-t)$, then each of the path probabilities can be decomposed into products of independent contributions from the realizations of each noise term. As shown in \cite{Note1}, these can be treated in the usual way \cite{Li2021, Cates2022, Nardini2017, Tauber2014}, leading to an expression for the informational entropy production in our system Eq.\,(\ref{eq:epr}) which is a sum of these individual contributions
\begin{equation}\label{eq:epr-am}
	\dot{\mathcal{S}} = -\lim_{\tau\rightarrow\infty} \frac{1}{D\tau}\int_0^\tau dt\int d\textbf{r}  \sum_{i=1}^N\left\langle \mu_i \dot{\phi}_i \right\rangle.
\end{equation}

Furthermore, if we decompose each chemical potential $\mu_i$ into equilibrium and nonequilibrium contributions $\mu_i =\mu_i^{\rm (eq)} + \mu_i^{\rm (neq)}$ and define the \textit{free-energy} functional, $\mathcal{F}[\{\phi_{i=\{1, \dots, N\}}\}]$ such that the equilibrium contribution is written as $\mu_i^{(\rm eq)}=\delta\mathcal{F}/\delta \phi_i$, then $\dot\phi \mu_i^{\rm (eq)}=\dot{\mathcal{F}}$ and Eq.\,(\ref{eq:epr-am}) simplifies to 
\begin{equation}\label{eq:final-epr}
\dot{\mathcal{S}} = -\lim_{\tau\rightarrow\infty} \frac{1}{D\tau}\int_0^\tau dt\int d\textbf{r}  \sum_{i=1}^N\left\langle \mu_i^{(\rm neq)}\dot{\phi}_i\right\rangle,
\end{equation}
provided that the free energy $\mathcal{F}$ is bounded in time \cite{Cates2022, Note1, Nardini2017}. As expected, Eq.~\eqref{eq:final-epr} shows that the dynamics are symmetric in time ($\dot{\mathcal{S}}=0$) in the absence of nonequilibrium contributions to the chemical potential. This constitutes our general result for scalar active mixtures and holds for any arbitrary nonequilibrium term provided we employ a suitable discretization scheme and the noise terms are independent \cite{Note1}. Note that we conveniently recover the result for the steady-state entropy production rate of \textit{Active Model B} by setting $N=1$ and substituting in the active term $\mu_{1}^{\rm (neq)}=\lambda|\nabla\phi_1|^2$ \cite{Nardini2017}.

\textit{Scalar active mixtures with nonreciprocal couplings. ---} To study the link between TRS breaking and nonreciprocity, we consider the nonreciprocal scalar field theory introduced in \cite{Saha2020}, which extends the classical Cahn-Hilliard model \cite{Cahn1958,Hohenberg1977} to include nonreciprocal linear couplings ({cross-diffusion}). As such, the field theory describes phase separation in a wide class of scalar active matter systems where reciprocal-symmetry breaking interactions appear at the continuum level \cite{Keller2006, Friedl2009, Zhao2018, Sweetlove2018}.

The governing equations for this field theory take the form \cite{Saha2020}:
\begin{equation}
	\dot{\phi}_i(\textbf{r}, t) = \nabla^2 \bigg[\frac{\delta\mathcal{F}}{\delta\phi_i}+ \sum_j \alpha_{ij} \phi_j(\textbf{r}, t)\bigg] + \nabla\cdot \mathbf{\Lambda}_i(\textbf{r}, t)
	\label{eq:nrft}
\end{equation}
where, as before, $i\in\{1, \dots , N\}$ and we have defined the global free-energy-like functional as 
\begin{equation}
\mathcal{F}[\{\phi_i\}] = \int d\textbf{r} \Big(\sum_{i} f_i(\phi_i) + \sum_{i<j}\kappa_{ij}\phi_i\phi_j\Big)
\end{equation} 
and $\{\alpha_{ij}\}_{i,j\in[1,N]}$ is a fully antisymmetric matrix. The functional includes two contributions: the first determines how each field evolves in isolation and the second describes the (reciprocal) enthalpic interactions between fields. Henceforth, we suppose that the free energy densities are of \textit{Model B} form: $f_i(\phi_i) = \chi_i{\phi_i^2}/{2} + {\phi_i^4}/12 + {\gamma_i}|\nabla\phi|^2/{2} $, where $\chi_i$ controls whether each field phase separates and $\gamma_i$ sets the effective surface tension when interfaces arise in the system \cite{Bray2002, Tjhung2018}. Note that in the absence of nonreciprocal couplings ($\alpha_{ij} \equiv 0$), the field theory described here is entirely equilibrium (or passive).

\textit{Breaking TRS through nonreciprocity. ---} Now we call upon Eq.\,(\ref{eq:final-epr}) to write an expression for the entropy production rate in our system with nonreciprocal couplings:
\begin{equation}\label{eq:nr-epr}
	\dot{\mathcal{S}} = -\lim_{\tau\rightarrow\infty} \frac{1}{D\tau}\int_0^\tau dt\int d\textbf{r}  \sum_{i=1}^N\sum_{j\ne i}\left\langle \alpha_{ij}\phi_j\dot{\phi}_i\right\rangle.
\end{equation}
Again, the case where $\alpha_{ij}=0$ for all $i$ and $j$ corresponds to purely reciprocal couplings and gives a zero steady-state rate of entropy production as expected. Note that while linear-order couplings (as the lowest order appearing in a gradient expansion) were studied in \cite{Saha2020} for the sake of simplicity, our first key result [Eq.\,(\ref{eq:final-epr})] holds for arbitrary nonreciprocal couplings and is thus valid beyond the class of systems described by Eq.\,(\ref{eq:nrft}) (see \cite{Saha2022} for a recent work on nonlinear nonreciprocity).
 
To illustrate our result, we confine our system to one spatial dimension on $[0, L)$ with periodic boundary conditions and set $N=2$. In this case, the strength of the nonreciprocal coupling can be controlled by a single parameter. Indeed, we can write the coupling coefficients as $\kappa_{12} = \kappa_{21} = \kappa$ and $\alpha_{12} = - \alpha_{21} = \alpha$. The resulting equations governing the dynamics of the two fields, which we denote $\phi_1(r, t)$ and $\phi_2(r, t)$, take the form
\begin{subequations}
\begin{align}
		\dot{\phi}_1(r, t) = \partial_r^2\mu^{\rm (eq)}_1(r, t) + \alpha \partial_r^2 \phi_2(r, t) + \partial_r \Lambda_1(r, t) \label{eq:1DNR1}\\
		\dot{\phi}_2(r, t) = \partial_r^2\mu^{\rm (eq)}_2(r, t) - \alpha \partial_r^2 \phi_1(r, t) + \partial_r \Lambda_2(r, t) \label{eq:1DNR2}
\end{align}\label{eq:1DNR}%
\end{subequations}
where the chemical potentials are again defined as the following functional derivatives of the free-energy-like functional $\mu^{\rm (eq)}_i = \frac{\delta\mathcal{F}}{\delta\phi_i}$ and $\Lambda_i(r, t)$ are zero-mean Gaussian white noise terms with diagonal covariance matrix $\Theta_{ij}(r-r', t-t')=2D\delta_{ij}\delta(r-r')\delta(t-t')$.  

From Eq.\,(\ref{eq:nr-epr}), we determine the expression for the steady-state rate of informational entropy production, $\dot{\mathcal{S}}$, for this binary system:
\begin{equation}\label{eq:1DNREP}
	\dot{\mathcal{S}} = - \lim_{\tau\rightarrow\infty} \frac{\alpha}{D\tau }\int_0^\tau dt \int_0^L dr \: \left\langle\phi_2\dot\phi_1 - \phi_1\dot\phi_2\right\rangle~,
\end{equation}
which vanishes at $\alpha=0$.

\begin{figure}
	\centering
	\includegraphics[width=0.44\textwidth]{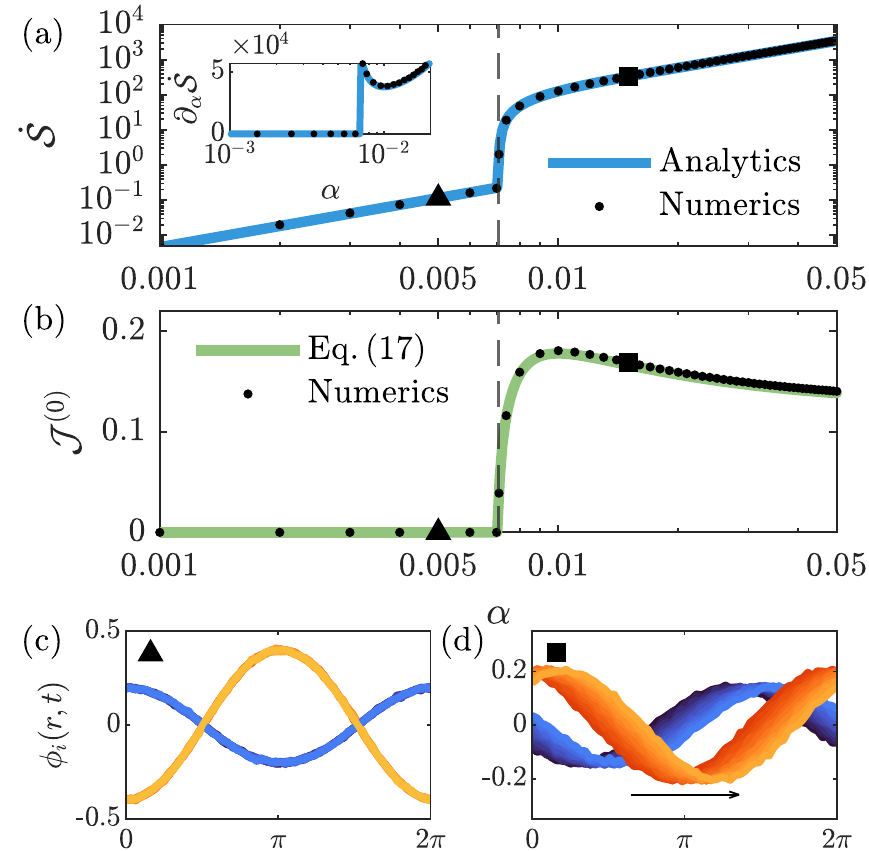}
		\caption{\textit{Non-trivial scaling of $\dot{\mathcal{S}}$ across ${\cal PT}$-breaking phase transition in a nonreciprocal system.} (a) For $D=10^{-7}$, we identify a non-trivial scaling in the entropy production rate in the binary system governed by Eq.\,(\ref{eq:1DNR}), with a discontinuous derivative (inset) at a critical value of the nonreciprocal coupling $\alpha=\alpha_c$.  The points are results of numerical simulations and the solid line is our analytic expression for the leading order contribution, stemming from Eqs.\,(\ref{eq:Sa_exp}) and (\ref{eq:Sb_exp}). (b) The discontinuity in the scaling of $\dot{\mathcal{S}}$ coincides with the breaking of ${\cal PT}$ symmetry, characterized by the non-zero value of the polar order parameter $\mathcal{J}^{(0)}$ for the corresponding deterministic equations (corresponding to $D=0$ in Eq.\,(\ref{eq:1DNR})). The critical nonreciprocal coupling corresponds to a transition between (c) static states and (d) dynamic states (travelling waves).}
	\label{fig:v_EP}
\end{figure}

We {first} explore the entropy production rate in this system numerically, quantifying how $\dot{\mathcal{S}}$ scales with the strength of the nonreciprocal coupling, set by $\alpha$. We work in the limit of  weak noise ({$D=10^{-7}$}) and further place ourselves in the case where (i) $\phi_1$ phase separates ($\chi_1<0$ and $\gamma_1>0$) and (ii) $\phi_2$ is purely diffusive. We also suppose that the two fields feel a weak (reciprocal) repulsion ($\kappa>0$) \cite{You2020}. We solve Eq.\,(\ref{eq:1DNR}) (see details of our numerical method in \cite{Note1}) then evaluate the integral in Eq.\,(\ref{eq:1DNREP}) numerically. As seen in Fig.\,\ref{fig:v_EP}(a), the entropy production rate initially scales as $\dot{\mathcal{S}}\propto \alpha^2$. At a critical value of the nonreciprocal coupling $\alpha=\alpha_c$, this scaling disappears and $\dot{\mathcal{S}}$ quickly increases continuously by {several orders} of magnitude. As $\alpha \gg \alpha_c$, we recover a quadratic scaling. 

To explain this non-trivial scaling in the entropy production rate, we explore in more details the dynamics of the system governed by Eq.\,(\ref{eq:1DNR}) (see also \cite{You2020}). In particular, we turn momentarily to the deterministic case, $D=0$, and observe that for $\alpha < \alpha_c$, the system reaches a {static} stationary state, where the two fields phase separate and exhibit demixing behavior [see Fig.\,\ref{fig:v_EP}(c)]. For $\alpha > \alpha_c$, the system instead displays a travelling wave solution and we observe the emergence of a non-zero global polar order $\mathcal{J}$ [see Fig.\,\ref{fig:v_EP}(b-d)], as defined by:
\begin{equation}\label{eq:J}
\mathcal{J}=\frac{1}{\tau}\int_0^\tau dt \int_0^L dr \big\langle\phi_2\partial_{r} \phi_1 - \phi_1\partial_{r} \phi_2\big
\rangle
\end{equation} 

In particular, it is clear that this order parameter becomes non-zero when the steady-state solution explicitly breaks the joint ${\cal PT} : r,t \mapsto -r,-t$ symmetry; said differently, ${\cal J}$ becomes non-zero when the interacting fields are out of phase in such a way that breaks ${\cal PT}$-symmetry. As discussed in \cite{You2020}, this transition to a dynamic phase is thus an example of ${\cal PT}$-symmetry breaking transition, where the asymmetric distribution of the fields leads to a imbalance of effective forces and thus persistent motion in the macroscopic dynamics (a so-called ``run-and-catch" mechanism \cite{Zhang2022, Soto2014, Saha2019, AgudoCanalejo2019, You2020}).

Note that, in principle, adding noise to the system will induce a non-zero rate of reversals in the travelling wave solutions. We work in the regime of weak noise strength so these reversals are rare events. The thermodynamic quantities derived below are valid for the dynamics between reversal events. 

\begin{figure}
	\centering
	\includegraphics[width=0.44\textwidth]{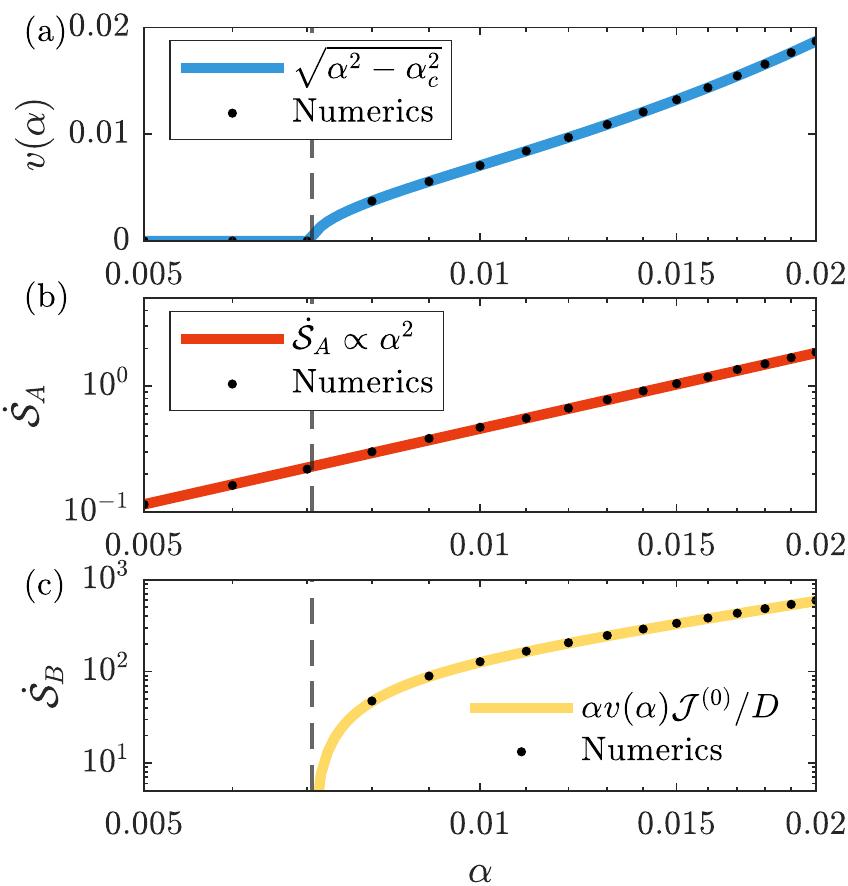}
		\caption{\textit{Evaluation of contributions to entropy production rate $\dot{\mathcal{S}}$ ---} (a) The transition to motion observed when the strength of the non-reciprocal coupling exceeds a critical value, $\alpha_c = \sqrt{\kappa^2+\chi_2^2}$. (b) The first contribution $\dot{\mathcal{S}}_A$ is evaluated numerically and is shown to scale quadratially with the non-reciprocal parameter $\alpha$. (c) We identify a second contribution to the entropy production rate stemming directly from the emergence of macroscopic dynamics as exhibited in (a), which we denote $\dot{\mathcal{S}}_B.$ }
	\label{fig:JOSB}
\end{figure}

\textit{Emergence of macroscopic dynamics and entropy production. ---} This second-order transition to motion controlled by $\alpha$ coincides with the non-trivial scaling of the entropy production rate as seen in Fig.\,1. We now formally connect these two phenomena by identifying the contribution of the macroscopic dynamics to the steady-state entropy production rate. To do this, we consider a change of variables to rewrite Eq.\,(\ref{eq:1DNREP}) in the comoving frame of reference. We let $v(\alpha)$ denote the velocity of the traveling wave solution [see Fig.\,\ref{fig:JOSB}(a)] and proceed to the transformation $(r', t')= (r - v(\alpha)t, t)$. We denote the fields $\phi_i$ in this new frame of reference by $\Phi_i$, such that the time derivatives in Eq.\,(\ref{eq:1DNREP}) now appear in the form 
$\dot{\phi}_i \rightarrow \dot{\Phi}_i - v(\alpha)\partial_{r'}\Phi_i$ and the entropy production rate now takes the form $\dot{\mathcal{S}} = \dot{\mathcal{S}}_{A}(\alpha) + \dot{\mathcal{S}}_{B}(\alpha)$ with
\begin{subequations}
\begin{align}
\dot{\mathcal{S}}_A &= \lim_{\tau\rightarrow\infty} \frac{\alpha}{D\tau}\int_0^\tau dt' \int_0^L dr'  \left\langle\Phi_1\dot\Phi_2 - \Phi_2\dot\Phi_1\right\rangle, \label{eq:S_AS_B1}\\
\dot{\mathcal{S}}_B &=\lim_{\tau\rightarrow\infty} \frac{\alpha v(\alpha)}{D\tau }\int_0^\tau dt' \int_0^L dr' \Big\langle\Phi_2\partial_{r'} \Phi_1 - \Phi_1\partial_{r'} \Phi_2\Big\rangle. \label{eq:S_AS_B2}
\end{align}\label{eq:S_AS_B}%
\end{subequations}

As seen in Fig.\,\ref{fig:JOSB}(b), we observe that $\dot{\mathcal{S}}_A \propto \alpha^2$ and independent of the travelling wave speed; it describes the entropy production due to the nonequilibrium suppression of fluctuations in a stationary phase-separated system, reminiscent of the entropy production observed in \textit{Active Model B} \cite{Wittkowski2014, Cates2022} which also scaled quadratically with the nonequilibrium contribution to the dynamics \cite{Nardini2017}. 

The second term $\dot{\mathcal{S}}_B$ captures the contribution of the macroscopic motion to the total entropy production rate, vanishing when $v(\alpha)=0$. Strikingly, the integral contribution to $\dot{\mathcal{S}}_B$ is exactly the global polar order parameter defined in Eq.\,(\ref{eq:J}), which implies $\dot{\mathcal{S}}_B=\alpha v(\alpha)\mathcal{J}/D$ and thus gives us a direct link between the macroscopic dynamics and informational entropy production, i.e. time-reversal symmetry breaking. 

\textit{Weak-noise expansion. ---} From Eq.\,(\ref{eq:final-epr}), we expect that signatures of TRS breaking are most striking when the noise is weak. In this regime, we can make progress towards evaluating the two contributions analytically by expanding each of the fields perturbatively around the deterministic solution ($D=0$):
\begin{align}
\Phi_i(r', t') = \Phi_i^{(0)}(r') &+ \sqrt{D}\Phi_i^{(1)}(r', t') \nonumber\\
    &+ D\Phi_i^{(2)}(r', t') + \mathcal{O}(D^{3/2}),
\label{eq:phi_exp}
\end{align}
Taking a time derivative of Eq.\,(\ref{eq:phi_exp}), one can derive governing equations for the dynamics of each field $\Phi_i^{(j)}$, which are independent of $D$ (see details in \cite{Note1}). The two distributions $\Phi_i^{(0)}$ are given by the deterministic solutions to Eq.\,(\ref{eq:1DNR}) in the frame of reference $(r', t')$. 

\begin{figure}
	\centering
	\includegraphics[width=0.42\textwidth]{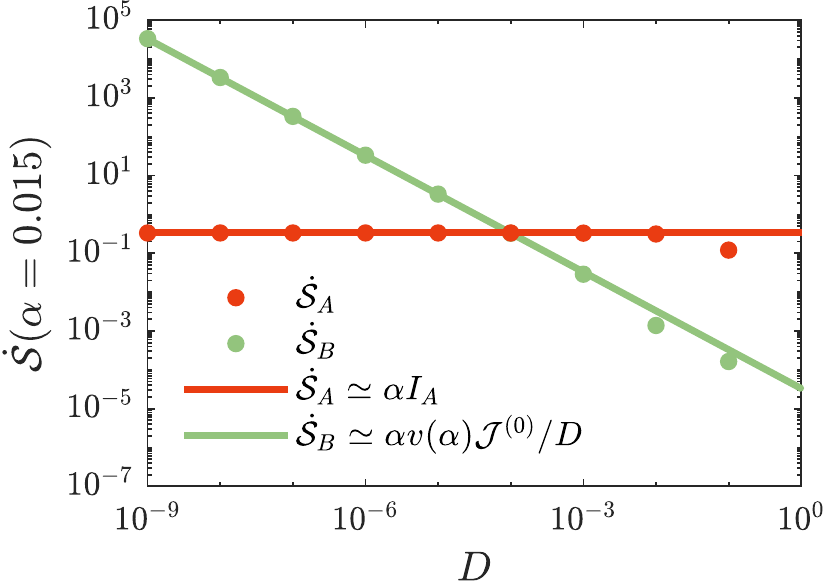}
		\caption{\textit{Scaling of $\dot{\mathcal{S}}_A$ and $\dot{\mathcal{S}}_B$ in weak-noise regime ---} We choose $\alpha>\alpha_c$ to ensure that both contributions to $\dot{\mathcal{S}}$ are non-zero and study their scaling with the diffusion coefficient $D$ in the weak noise regime. We confirm through the numerical simulations the analytic results obtained through the one-mode approximation: $\dot{\mathcal{S}}_A\propto D^0$ as shown in Eq.\,(\ref{eq:Sa_exp}) (where the pre-factor is determined from fitting the data) and $\dot{\mathcal{S}}_B\propto D^{-1}$ as shown in Eq.\,(\ref{eq:Sb_exp}).}
	\label{fig:Dscale}
\end{figure}

Substituting Eq.\,(\ref{eq:phi_exp}) into Eq.\,(\ref{eq:S_AS_B1}), the terms of order $\mathcal{O}(D^{-1})$ and $\mathcal{O}(D^{-1/2})$ disappear in the expansion for $\dot{\mathcal{S}}_A$ at steady-state (see discussion in \cite{Note1}), thus to leading order this contribution is written as
\begin{equation}
    \dot{\mathcal{S}}_A =  \alpha I_A(\alpha)+\mathcal{O}({D}^{1/2})\\
    \label{eq:Sa_exp}
\end{equation}
with
\begin{equation}
    I_A(\alpha)=\lim_{\tau\rightarrow\infty}\frac{1}{\tau}\int_0^\tau dt \int_0^L dr\big\langle\Phi_1^{(1)}\dot{\Phi}_2^{(1)} - \Phi_2^{(1)}\dot{\Phi}_1^{(1)}\big\rangle.
\end{equation}
Therefore, $\dot{\mathcal{S}}_A\propto D^0$ in the small $D$ regime, which we confirm against numerical results in Fig.\,\ref{fig:Dscale}. To obtain a closed analytic expression, we require the form of $\Phi_i^{(1)}$, but the governing equations for these fields (derived in \cite{Note1}) do not generally admit analytic solutions.

To leading order, we further write that
\begin{equation}
    \dot{\mathcal{S}}_B = \frac{\alpha v(\alpha)}{D}\mathcal{J}^{(0)}+\mathcal{O}(D^{-1/2}),
    \label{eq:Sb_exp}
\end{equation}
where $\mathcal{J}_0$ denotes the global polar order parameter evaluated in the deterministic limit
\begin{equation}
{\cal J}^{(0)} = \lim_{\tau\rightarrow\infty} \frac{1}{\tau } \int_0^{\tau}dt' \int_0^{L} dr'\: \bigg( \Phi_2^{(0)} \partial_{r'}\Phi_1^{(0)} -\Phi_1^{(0)} \partial_{r'}\Phi_2^{(0)}\bigg),
\end{equation}
Note that we have discarded the average over noise realizations as $\Phi_{1}^{(0)}$ and $\Phi_2^{(0)}$ are --- by construction of the expansion in Eq.\,(\ref{eq:phi_exp}) --- solutions to the deterministic governing equations and thus noise-independent.

It follows that we only require analytic expressions for the deterministic solutions to Eq.\,(\ref{eq:1DNR}) to write the leading order contribution to $\dot{\mathcal{S}}_B$. In the parameter regime explored in \cite{You2020}, namely when $|\chi_1/\gamma_1| \gtrsim (2\pi/L)^2$, one can argue that only the lowest allowed wavenumber mode $q=2\pi/L$ is linearly unstable and it is thus expected to dominate the structure of the stationary distributions in a Fourier series expansion (see \cite{Note1} for details), which is not expected in general \cite{FrohoffHulsmann2021}. As such, we can proceed to a one-mode approximation for the stationary distributions, in effect writing $\Phi_i^{(0)}(r')\propto \cos(2\pi r'/L-\theta_i)$, where $\theta_1$ can be set to zero by translational invariance and $\theta_2$ sets the difference in phase of the distributions.

As shown in \cite{Note1}, in the case where $L=2\pi$, we find under this one-mode approximation that
\begin{equation}
	\mathcal{J}^{(0)} = \frac{8\pi v(\alpha)(\chi_1+\chi_2+\gamma_1)}{(\kappa - \alpha)},
\end{equation}
giving us a closed-form analytic expression for $\dot{\mathcal{S}}_B$ to leading order. Overall, we conclude that $\dot{\mathcal{S}}_B \propto D^{-1}$ in the small D regime which agrees with the numerical results of Fig.\,\ref{fig:Dscale}. Interestingly, the parameter space corresponding to the existence of a travelling wave solution is a subset of the space required for $\dot{\mathcal{S}}_B>0$, guaranteeing that the second law of thermodynamics is strictly satisfied.  Strikingly, we also recover an expression relating the entropy production rate, the travelling wave velocity and diffusion coefficient: $\dot{\mathcal{S}}_B\propto v^2(\alpha)/D$, which becomes more accurate when $\alpha\gg \kappa$. This exact scaling relation appears when studying the entropy production rate for a self-propelled particle with propulsion speed $v$ and diffusion coefficient $D$ \cite{Seifert2012,Cocconi2020}. Here, we show that this scaling captures a deeper connection between motion and time-reversal symmetry-breaking in active systems. 

\textit{Discussion and outlook. ---} In this Letter, we have studied the time-reversal symmetry breaking implications of introducing nonreciprocal couplings to passive field theories by evaluating the rate of informational entropy production. We first derived general analytical results for a mixture of conserved active fields assuming additive and independently distributed noise terms. We then employed this result to quantify TRS breaking in a binary mixture of active fields in the presence of nonreciprocal interactions. In this case, we showed that the informational entropy production rate has a non-trivial scaling with the strength of the nonreciprocal interactions due to the underlying emergence of macroscopic dynamics. Such discontinuities in the scaling of the entropy production rate have also been observed in self-propelled microscopic systems at the transition to collective motion \cite{Yu2022,Ferretti2022}. This suggests a more general link between the thermodynamic properties of nonequilibrium systems and the breaking of ${\cal PT}$-symmetry in transitions from a static to a dynamic phase \footnote{We note that interestingly a recent study obtained independently results consistent with the theory we present in this manuscript, see \cite{Suchanek2023}}. To elucidate this, we derive here an analytic expression for the leading-order contribution to the entropy production rate in the weak-noise regime and show the predominance of the contribution associated with the emergence of motion at the macroscale.

Future work will address extending these methods to a broader class of active mixtures, including in the presence of non-conservative dynamics. This could include active systems with Markovian switching which models, for example, chemical reactions \cite{Zwicker2014,Zwicker2015,Weber2019} or microscopic changes of state \cite{Alston2022a, Alston2022b, MonchoJorda2020, Bley2021}. We also wish to extend our methodology to systems with correlated or multiplicative noise; the corresponding problem for isolated active fields has seen some recent attention \cite{Markovich2021, dePirey2022, Cugliandolo2017, Cugliandolo2019}. Constructing a framework with which to analyze the nonequilibrium thermodynamic properties of these active and living systems is a major challenge with applications across disciplines and one which we expect to draw considerable attention in the immediate future.

\begin{acknowledgments}
HA was supported by a Roth PhD scholarship funded by the Department of Mathematics at Imperial College London. LC acknowledges support from the Francis Crick Institute, which receives its core funding from Cancer Research UK, the UK Medical Research Council, and the Wellcome Trust (FC001317). 
\end{acknowledgments}


%

\end{document}


\title{Supplemental Material for ``Irreversibility across a nonreciprocal $\mathcal{PT}$-symmetry-breaking phase transition"}

	\author{Henry Alston}
	\affiliation{Department of Mathematics, Imperial College London, South Kensington, London SW7 2AZ, United Kingdom}

	\author{Luca Cocconi}
	\affiliation{Department of Mathematics, Imperial College London, South Kensington, London SW7 2AZ, United Kingdom}
	\affiliation{The Francis Crick Institute, London NW1 1AT, United Kingdom}
	\affiliation{Max Planck Institute for Dynamics and Self-Organization (MPIDS), 37077 G\"{o}ttingen, Germany}
			
	\author{Thibault Bertrand}%
	\email{t.bertrand@imperial.ac.uk}
	\affiliation{Department of Mathematics, Imperial College London, South Kensington, London SW7 2AZ, United Kingdom}

\renewcommand{\theequation}{S\arabic{equation}}
\renewcommand{\thefigure}{S\arabic{figure}}

\maketitle

\hrule
\tableofcontents
\vspace{2em}
\hrule

\section{Entropy Production for a Generic Active Mixture in Continuous Space} 
\label{sec:EPRcontspace}
Here, we give the full derivation of the general result for the informational entropy production rate $\dot{\mathcal{S}}$ given in the main text, following a now standard procedure for nonequilibrium stochastic field theories in continuous space \cite{Tauber2014, Nardini2017, Li2021}. 
Consider a mixture of $N$ conserved components described by continuous scalar fields with dynamics governed by the set of Langevin equations
\begin{equation}\label{eq:eom}
\dot{\phi}_i(\textbf{r}, t) = \nabla^2 \mu_i(\textbf{r}, t) + \nabla \cdot \bm{\Lambda}_i(\textbf{r}, t),
\end{equation}
coupled via the chemical potentials $\mu_i(\textbf{r}, t)$, which can include both active and passive contributions. Here, $\mathbf{\Lambda}$ is a vector of zero-mean Gaussian white noise terms with diagonal correlation matrix 
\begin{equation}\label{eq:cont_noise_corr}
\langle \bm{\Lambda}_i(\textbf{r}, t)\bm{\Lambda}_j(\textbf{r}', t')\rangle = 2D\delta_{ij}\delta(\mathbf{r}'-\mathbf{r})\delta(t'-t). 
\end{equation}

We measure the extent to which the joint dynamics of the $N$ fields breaks time-reversal symmetry by evaluating the Kullback-Leibler divergence per unit time of the ensemble of forward paths and their time-reversed counterparts \cite{Seifert2012,Cocconi2020},  
\begin{equation}\label{eq:epr}
    \dot{\mathcal{S}} = \lim_{\tau\rightarrow\infty} \frac{1}{\tau} \bigg\langle \log \frac{\mathbb{P}_F[\{\phi_i\}_{t=0}^{t=\tau}]}{\mathbb{P}_B[\{\phi_i\}_{t=0}^{t=\tau}]}\bigg\rangle, 
\end{equation}
where the $\mathbb{P}_F[\cdot]$ and $\mathbb{P}_B[\cdot]$ denote the path probability density for the forward and time-reversed paths respectively while $\langle \cdot \rangle$ indicates averaging over realizations of the noise $\mathbf\Lambda$. For ergodic processes, this ensemble average can be omitted without affecting the result.
The quantity calculated in Eq.\,\eqref{eq:epr} can be understood as the steady-state rate of informational entropy produced in the system of coarse-grained fields $\{\phi_i\}$, not to be confused with the thermodynamic entropy production of the underlying microscopic dynamics. 

The set of Langevin equations \eqref{eq:eom} defines a map between any given realization of the $N$ noise terms $\bm{\Lambda}_i$ and a particular trajectory in the space of the $N$ fields.
Thus, the path probability for a $N$-field trajectory $\{\phi_i\}_{t=0}^{t=\tau}$ can be written entirely in terms of the noise realizations as 
\begin{equation}
\mathbb{P}_F[\{\phi_i\}_{t=0}^{t=\tau}] = \mathcal{J}[\{\phi_i\}_{t=0}^{t=\tau}] \mathbb{P}_F[\{{\bm{\Lambda}}_i\}_{t=0}^{t=\tau}] = \mathcal{J}[\{\phi_i\}_{t=0}^{t=\tau}] \prod_{i=1}^N \mathbb{P}_F[({\bm{\Lambda}}_i)_{t=0}^{t=\tau}]
\end{equation}
where $\mathcal{J}[\phi] = \mathcal{D}[\bm{\Lambda}]/\mathcal{D}[\phi]$ denotes the Jacobian of the field transformation, which in general depends on the choice of time discretization for the stochastic dynamics \cite{Tauber2014}. In the last equality, we have used the fact that the $N$ noise terms are statistically independent by Eq.\,\eqref{eq:cont_noise_corr}. The path probability densities for each noise component are Gaussian,
\begin{equation}\label{eq:path-noise}
\mathbb{P}_F[\bm{\Lambda}_i(\textbf{r}, t)_{t=0}^{t=\tau}] \propto \exp\bigg[-\frac{1}{2}\int_0^\tau dt\int d\textbf{r} \: |\bm{\Lambda}_i(\textbf{r}, t)|^2\bigg]
\end{equation}
which again follows from the definition of the noise. After re-arranging Eq.\,(\ref{eq:eom}) to write each noise component ${\Lambda}_i$ in terms of the physical field $\phi_i$ and the corresponding chemical potential $\mu_i$, we substitute into the right hand side of Eq.\,(\ref{eq:path-noise}) and finally integrate by parts once to obtain the forward path probability for the dynamics of the field $\phi_i$
\begin{equation}\begin{aligned}\label{eq:path-forward}
    \mathbb{P}_F[\phi_i(\textbf{r}, t)_{t=0}^{t=\tau}] \propto \mathcal{J}[\{\phi_i\}_{t=0}^{t=\tau}] \exp\bigg[-\frac{1}{4D} \int_0^\tau dt\int d\textbf{r} \ (\dot{\phi}_i - \nabla^2 \mu_i) \circ \nabla^{-2}(\dot{\phi}_i-\nabla^2\mu_i)\bigg]
\end{aligned}\end{equation}
where $\nabla^{-2}$ is the inverse Laplacian operator (with $\nabla^{-2} \nabla^2 = 1$, defined up to a choice of gauge as discussed in the next section) and $\circ$ denotes the Stratonovich product. 
Similarly, using $\mathbb{P}_B[\{\psi(t)\}_{t=0}^{t=\tau}] = \mathbb{P}_F[\{ \psi(t'=\tau-t)\}_{t=0}^{t=\tau}]$ together with the fact that the chemical potentials are even under time-reversal, we obtain the backward path probability as
\begin{equation}
    \mathbb{P}_B[\phi_i(\textbf{r}, t)_{t=0}^{t=\tau}] \propto \mathcal{J}[\{\phi_i\}_{t=0}^{t=\tau}] \exp\bigg[-\frac{1}{4D} \int_0^\tau dt\int d\textbf{r} \ (\dot{\phi}_i + \nabla^2 \mu_i) \circ \nabla^{-2}(\dot{\phi}_i+\nabla^2\mu_i)\bigg]~.
\end{equation}
The invariance of the Jacobian $\mathcal{J}$ under time reversal, and thus its cancellation upon taking ratios of forward and time-reversed path probabilities densities, is specific to the Stratonovich mid-point discretization, which motivates our choice of the latter.

Returning to Eq.\,(\ref{eq:epr}) for the informational entropy production rate $\dot{\mathcal{S}}$, we substitute the expressions for the respective path probabilities to obtain
\begin{equation}\label{eq:epr-am}
    \dot{\mathcal{S}} = -\lim_{\tau\rightarrow\infty} \frac{1}{D\tau}\int_0^\tau dt\int d\textbf{r}  \sum_{i=1}^N\langle \mu_i \circ \dot{\phi}_i \rangle.
\end{equation}
Finally, we decompose the chemical potential into its equilibrium and nonequilibrium contributions as $\mu_i=\mu_i^{(\rm eq)} + \mu_i^{(\rm neq)}$, where the contribution $\mu_i^{(\rm eq)}$ contains all terms that can be subsumed into a functional derivative of a global free-energy-like functional, $\mathcal{F}[\{\phi_i\}]$, taking the form $\mu_i^{(\rm eq)}={\delta\mathcal{F}}/{\delta \phi_i}$. Substituting into Eq.\,(\ref{eq:epr-am}), we have
\begin{equation}
    \dot{\mathcal{S}} = -\lim_{\tau\rightarrow\infty} \frac{1}{D\tau}\int_0^\tau dt\int d\textbf{r}  \sum_{i=1}^N\bigg[\bigg\langle \frac{\delta\mathcal{F}}{\delta \phi_i} \circ \dot{\phi}_i \bigg\rangle + \langle \mu_i^{(\rm neq)}\circ \dot{\phi}_i\rangle\bigg].
\end{equation}
The first term in the integral amounts to a total derivative of the free energy with respect to time and thus results in a contribution to the entropy production rate of the form $(\mathcal{F}[t=0]-\mathcal{F}[t=\tau])/D\tau$ which goes to zero in the limit $\tau\rightarrow\infty$, provided that the free energy is bounded from below. We conclude on the final form of the entropy production rate as
\begin{equation}\label{eq:final-epr}
    \dot{\mathcal{S}} = -\lim_{\tau\rightarrow\infty} \frac{1}{D\tau}\int_0^\tau dt\int d\textbf{r}  \sum_{i=1}^N\langle \mu_i^{(\rm neq)}(\textbf{r},t) \circ \dot{\phi}_i(\textbf{r},t)\rangle~,
\end{equation}
which is the form used in the main text. Since $\mu_i$ is typically a quasi-local function of the fields $\phi_j$ through their derivatives and powers thereof, Eq.\,\eqref{eq:final-epr} allows us to express the entropy production in terms of a set of equal-position equal-time correlation functions. Unfortunately, it is a common features of dynamical field theories that these correlations are  divergent when evaluated in the continuum \cite{Tauber2014}. In the following section, we address this problem by means of a lattice regularization.
    
\section{Entropy Production for a Generic Active Mixture on a Lattice} 

As anticipated at the end of the previous section, a na{\"i}ve attempt at evaluating the informational entropy production rate Eq.\,\eqref{eq:epr} for the dynamics in Eq.\,(\ref{eq:eom}) in continuous space leads to problems in the form of UV divergent contributions, as well as the appearance of ill-defined expressions involving the square of a Dirac delta function for specific choices of the chemical potential \cite{Nardini2017}. This is not completely surprising, since the flat power spectrum of the white noise \eqref{eq:cont_noise_corr} is of course unphysical.

Such problems can be avoided by working instead in discretized space, i.e.\ by mapping the continuum dynamics to those on a regular lattice, whereby the use of a finite lattice spacing $h$ effectively imposes a UV cutoff. This also ensures that the gradient noise term in Eq.\,(\ref{eq:eom}) is well-defined. In the same spirit, one might expect that ``softer'' regularization approaches, such as an exponential suppression of the noise spectrum at high momenta and frequencies (which are known to regularize the UV divergence of the coincident field correlation), would prove equally effective.
Remarkably, we show in Sec.\,\ref{sec:regUV} below that this alternative approach fails to regularize the UV divergence of the entropy production rate.

\subsection{Definition of Lattice Model}

We consider the dynamics of Eq.\,(\ref{eq:eom}) in one dimension mapped to a regular lattice with spacing $h$ and total length $L$ with periodic boundary conditions. We denote by $\phi_{j,k}$ the value of the $k$-th field at site $j$ on the lattice, where there are $N$ fields in total and $M=L/h$ sites. The same indexing rules are used for the $N$ chemical potentials and noise terms. We define the discrete gradient operator $\nabla_d$ and the discrete Laplacian $\nabla^2_d$ to be of centered finite difference form,
\begin{equation}\label{eq:grad_dis}
    \nabla_d \phi_{j, k} \equiv \delta_{k,k'} \nabla_{jj'} \phi_{j', k'} = \frac{\phi_{j+1, k}-\phi_{j-1, k}}{2 h}, \quad \nabla^2_d \phi_{j, k} \equiv  \delta_{k,k'} \nabla^2_{jj'} \phi_{j', k'} = \frac{\phi_{j+2, k}+\phi_{j-2, k}-2\phi_{j, k}}{4h^2}
\end{equation}
ensuring that detailed balance is satisfied in the absence of active contributions to the chemical potentials $\mu_{j, k}$ \cite{Nardini2017}.
Replacing the gradient and Laplace operators in Eq.\,(\ref{eq:eom}) with their discretized counterparts, we obtain 
\begin{equation}\label{eq:eom_SI_dis}
    \dot{\phi}_{j, k} = \frac{\mu_{j+2, k}+\mu_{j-2, k}-2\mu_{j, k}}{4h^2} + \sqrt{2D}\frac{(\Lambda_{j+1, k} - \Lambda_{j-1, k})}{2h}~.
\end{equation}
Finally, the correlator for the noise $\Lambda_{j,k}$ on the lattice is defined such that in the continuum limit $h\to 0$, we recover the Dirac delta correlator in continuous space as given in Eq.\,(\ref{eq:cont_noise_corr}). A suitable choice is 
 \begin{equation}\label{eq:disc_noise_corr}
     \langle \Lambda_{j, k}(t)\Lambda_{j', k'}(t')\rangle = \frac{\delta_{jj'}\delta_{kk'}\delta(t-t')}{h}~,
 \end{equation}
where $\delta_{\alpha\beta}$ denotes the Kronecker delta. 

The discrete Fourier transforms of the fields and their inverses are defined via
\begin{equation}\label{eq:dft}
    \phi_{j, k} = \frac{1}{M}\sum_{m=0}^{M-1} e^{2\pi i j m/M}\tilde{\phi}_{m, k}\quad\text{and}\quad  \tilde{\phi}_{m, k} = \sum_{j=0}^{M-1} e^{-2\pi i j m/M}\phi_{j, k}.
\end{equation} 
Based on Eq.\,\eqref{eq:dft}, discrete derivatives on the lattice amount to a trigonometric factor in momentum space
\begin{equation}
    \sum_{j=0}^{M-1}e^{-2\pi i j m/M} \big[\nabla_d\phi_{j, k}\big] = \frac{i\sin(2\pi m/M)}{h}\tilde{\phi}_{m, k}, \quad \sum_{j=0}^{M-1}e^{-2\pi i j m/M} \big[\nabla^2_d\phi_{j, k}\big] = \frac{-\sin^2(2\pi m/M)}{h^2}\tilde{\phi}_{m, k}~.
\end{equation}
Substituting into Eq.\,\eqref{eq:eom_SI_dis}, it follows that the governing equations for the Fourier modes take the form
\begin{equation}\label{eq:eom_dfs}
    \dot{\tilde{\phi}}_{m, k} = \frac{-\sin^2(2 \pi m/M)}{h^2} \tilde{\mu}_{m, k} + \frac{i\sqrt{2D}\sin(2\pi m/M)}{h}\tilde{\Lambda}_{m, k}~.
\end{equation}
where we denote by $\tilde{\mu}_{m, k}$ and $\tilde{\Lambda}_{m, k}$ the discrete Fourier transforms of the chemical potentials and noise terms respectively. 
Finally, using Eq.\,\eqref{eq:disc_noise_corr}, the noise correlator in Fourier space reads
\begin{equation}
    \langle \tilde{\Lambda}_{m, k}(t')\tilde{\Lambda}_{m', k'}(t) \rangle = \frac{M}{h} \delta_{kk'}\delta_{m'+m, 0}\delta(t'-t)~.
\end{equation}
While the conservative dynamics Eq.\,\eqref{eq:eom_dfs} are well defined for $m=0$, it is clear that these are trivial since each homogeneous mode $\tilde{\phi}_{0,k}$ is a constant. Equivalently, the homogeneous mode of the noise, $\tilde{\Lambda}_{0,k}$, does not couple to the field dynamics and we can therefore restrict the definition of the noise to the finite wavenumber modes $m\geq 1$, which turns out to be convenient in the following. 

\subsection{Onsager-Machlup and Informational Entropy Production Rate} 
\label{s:epr_fourier}
We now write the path probabilities as a function of the Fourier transformed noise terms by taking a discrete Fourier transform of the real-space Onsager-Machlup functional,
\begin{equation}\label{eq:fpp}
    P_F[\{\tilde{{\Lambda}}_{m, k}\}] \propto \exp{\bigg[-\frac{h}{2M}\sum_{k=1}^N\sum_{m=1}^{M-1}\tilde{\Lambda}_{m,k} \tilde{\Lambda}_{-m, k}\bigg]}~,
\end{equation}
where we have excluded the $m=0$ term in the sum over momenta on the basis that the dynamics are entirely independent of the realization of the noise $\tilde{\Lambda}_0$, as discussed above. Similarly to what was done for the continuum case, we now rearrange Eq.\,(\ref{eq:eom_dfs}) to write the noise term as a function of the fields, then substitute these expressions into the Eq.\,(\ref{eq:fpp}) to find the appropriate weighting for the forward and backward paths,
\begin{small}
\begin{subequations}
\begin{align}
    &P_F[\{\tilde{\phi}_{m, k}\}] \propto \exp{\bigg[-\frac{h}{2M}\sum_{m=1}^{M-1} \frac{h^2}{2D\sin^2(2m\pi/M)}\bigg(\dot{\tilde{\phi}}_{m, k} + \frac{\sin^2(2\pi m/M)\tilde{\mu}_{m, k}}{h^2}\bigg)\circ \bigg( \dot{\tilde{\phi}}_{-m, k} + \frac{\sin^2(2\pi m/M)\tilde{\mu}_{-m, k}}{h^2}\bigg)\bigg]}, \\
    &P_B[\{\tilde{\phi}_{m, k}\}] \propto \exp{\bigg[-\frac{h}{2M}\sum_{m=1}^{M-1} \frac{h^2}{2D\sin^2(2m\pi/M)}\bigg(\dot{\tilde{\phi}}_{m, k} - \frac{\sin^2(2\pi m/M)\tilde{\mu}_{m, k}}{h^2}\bigg)\circ \bigg( \dot{\tilde{\phi}}_{-m, k} - \frac{\sin^2(2\pi m/M)\tilde{\mu}_{-m, k}}{h^2}\bigg)\bigg]}.
    \end{align}
\label{eq:fpp2}
\end{subequations}
\end{small}

Using Eq.\,\eqref{eq:epr}, we can finally obtain an expression for the informational entropy production rate for the discretized dynamics of the form
\begin{align}
    \dot{\mathcal{S}}_{\rm disc.} &=-\lim_{\tau\rightarrow\infty}\frac{1}{D\tau}\sum_{k=1}^N\int_0^{\tau}dt\: \sum_{m=1}^{M-1}\frac{h}{2M}\big\langle\dot{\tilde{\phi}}_{-m, k}\circ \tilde{\mu}_{m, k} + \dot{\tilde{\phi}}_{m, k} \circ\tilde{\mu}_{-m, k}\big\rangle \nonumber \\
    &= -\lim_{\tau\rightarrow\infty}\frac{h}{MD\tau}\sum_{k=1}^N\int_0^{\tau}dt\: \sum_{m=1}^{M-1} \big\langle\dot{\tilde{\phi}}_{m,k} \circ \tilde{\mu}_{-m, k}\big\rangle
\end{align}
where we identify the sum over $m$ in the final integral as a convolution in Fourier space (up to the missing $m=0$ term). Finally, since $\partial_t \tilde{\phi}_{0,k}=0$ and $\tilde{\mu}_{0, k}$ is finite, we are free to add back the $m=0$ summand and use the definitions in Eq.\,(\ref{eq:dft}) to write this convolution as a product in real space as
\begin{equation}
    \sum_{m=0}^{M-1} \dot{\tilde{\phi}}_{m, k}\circ\tilde{\mu}_{-m, k} = M\sum_{j=0}^{M-1}\dot{\phi}_{j, k}\circ\mu_{j,k}~,
\end{equation}
whereby the informational entropy production rate for the discretized problem written in terms of real-space functions takes the form
\begin{equation}\label{eq:epr_disc}
    \dot{\mathcal{S}}_{\rm disc.} 
    =-\lim_{\tau\rightarrow\infty}\frac{h}{D\tau}\sum_{k=1}^N\sum_{j=0}^{M-1}\int_0^{\tau}dt\:  \big\langle \dot{\phi}_{j, k}\circ{\mu}_{j, k}\big\rangle 
    = -\lim_{\tau\rightarrow\infty}\frac{h}{D\tau}\sum_{k=1}^N\sum_{j=0}^{M-1}\int_0^{\tau}dt\:  \big\langle \dot{\phi}_{j, k}\circ{\mu}^{\rm (neq)}_{j, k}\big\rangle ~.
\end{equation}
Of course, Eq.\,\eqref{eq:epr_disc} converges to the result for continuous space, Eq.\,(\ref{eq:final-epr}), when we the continuum limit $h\rightarrow 0$ is taken. Indeed, this follows from the fact that
\begin{equation}
    \lim_{h\rightarrow 0}\bigg(h \sum_{j=0}^{M-1}\big[\dots\big]\bigg) \equiv \int_0^{Mh = L}dx\big[\dots\big] ,
\end{equation}
whereby $\lim_{h\rightarrow 0}\dot{\mathcal{S}}_{\rm disc.} = \dot{\mathcal{S}}$.

\subsection{Numerical Integration of Lattice Model}

We give here details of the numerical simulation of the lattice dynamics in Eq.\,(\ref{eq:eom_SI_dis}) that were used to test our analytical prediction. Eq.\,(\ref{eq:eom_SI_dis}) is integrated in time using a standard Euler-Maruyama method with explicit timestep $\Delta t=10^{-4}$ following the It\^{o} convention for the time discretization for the sake of convenience \cite{Kloeden2002} . Unless specified otherwise, the parameter values used in the simulations shown here are chosen to match those of Fig.\,1 in \cite{You2020}, namely $M=64,\,L=2\pi, \,\gamma_1=0.04, \,\gamma_2=0,\,\chi_1=-0.05,\,\chi_2=0.005\,\text{and}\,\kappa=0.005.$ We also set the homogeneous mode of each field equal to $0$. For simulations where the number of sites is varied, smaller timesteps are used, down to $2\times10^{-5}$ for $M=96$. For each realization of the dynamics, we initialize the system in a solution of the deterministic problem (which we rederive in Sec.\,\ref{s:lowD} below for completeness), oriented so that the travelling wave solution, should it exist, is propagating to the right. We then evolve the solution at finite noise strength $D>0$, averaging the relevant observables over $5 \times 10^3$ realizations of the noise as instructed by the $\langle \cdot \rangle$ brackets in Eq.\,(\ref{eq:epr_disc}). Each sample trajectory is of duration $\tau=10$ time units, having allowed the system to relax for an additional 10 time units after initialization to avoid biases associated with the choice of initial conditions. We deemed the 10 time unit period to be sufficient by comparing the results against longer periods, up to 100 time units.

\subsection{On Evaluation of the Stratonovich Integral (\ref{eq:epr_disc})}

For the particular case of a scalar active mixture with nonreciprocal linear couplings studied in this work, the numerical evaluation of Eq.\,(\ref{eq:epr_disc}) for the informational entropy production is very much simplified by converting the stochastic integral appearing in the latter from a Stratonovich to a It\^{o} discretization scheme. To do so, we first substitute the particular form for the nonequilibrium part of the chemical potential $\mu^{\rm (neq)}_{j,k} = \sum_{\ell \neq k}\alpha_{k\ell} \phi_{j,\ell}$ into (\ref{eq:epr_disc}) to obtain 
\begin{align}
    \dot{\mathcal{S}}_{\rm disc.} 
    &= -\lim_{\tau\rightarrow\infty}\frac{h}{D\tau}\sum_{k=1}^N\sum_{j=0}^{M-1} \sum_{\ell\neq k} \int_0^{\tau}dt\:  \alpha_{k\ell} \big\langle \dot{\phi}_{j, k}\circ \phi_{j,\ell} \big\rangle \label{eq:conv0}\\
    &= -\lim_{\tau\rightarrow\infty}\frac{h}{D\tau}\sum_{k=1}^N\sum_{j=0}^{M-1} \sum_{\ell\neq k} \int_0^{\tau}dt\:  \alpha_{k\ell} \left[ \big\langle (\nabla_d^2 \mu_{j,k} + \sqrt{2D} \nabla_d \Lambda_{j,k}) \phi_{j,\ell} \big\rangle + \langle \Xi_{j,(k,\ell)} \rangle\right] \label{eq:conv1}\\
    &= -\lim_{\tau\rightarrow\infty}\frac{h}{D\tau}\sum_{k=1}^N\sum_{j=0}^{M-1} \sum_{\ell\neq k} \int_0^{\tau}dt\:  \alpha_{k\ell} \left[ \big\langle \phi_{j,\ell} \nabla_d^2 \mu_{j,k} \big\rangle + \langle \Xi_{j,(k,\ell)} \rangle \right] \label{eq:conv2}
\end{align}
where the stochastic integrals \eqref{eq:conv1} and \eqref{eq:conv2} are now of the It\^{o} type, while in going from \eqref{eq:conv1} to \eqref{eq:conv2} we have used $\langle \phi_{j,k} \Lambda_{j',k'} \rangle=0$ for all $j',k'$ on the basis of the non-anticipating property of the It\^{o} integral. The conversion factor $\Xi_{j,(k,\ell)}$ is given by \cite{Cugliandolo2017,dePirey2022} 
\begin{equation}
    \Xi_{j,(k,\ell)} = \frac{1}{2} \sum_{j',\ell'} \frac{\partial \phi_{j,\ell}}{\partial \phi_{j', \ell'}} \delta_{\ell' k} \nabla^2_{jj'} = \sum_{j'} \frac{1}{2} \delta_{jj'} \delta_{\ell k} \nabla^2_{jj'} ~,
\end{equation}
thus vanishing for $\ell\neq k$, i.e.\ for all terms in Eq.\,\eqref{eq:conv2}. We are finally left with
\begin{equation}\label{eq:ito_int}
    \dot{\mathcal{S}}_{\rm disc.} = -\lim_{\tau\rightarrow\infty}\frac{h}{D\tau}\sum_{k=1}^N\sum_{j=0}^{M-1} \sum_{\ell\neq k} \int_0^{\tau}dt\:  \alpha_{k\ell} \big\langle \phi_{j,\ell} \nabla_d^2 \mu_{j,k} \big\rangle~.
\end{equation}
Eq.\,\eqref{eq:conv2} with $\Xi_{j,(k,\ell)}=0$, which amounts to evaluating \eqref{eq:conv0} in the It\^{o} discretization, is the expression we actually compute in our numerical simulations.

\subsection{Divergence of Entropy Production Rate in Continuum Limit}
\begin{figure}
    \centering
    \includegraphics{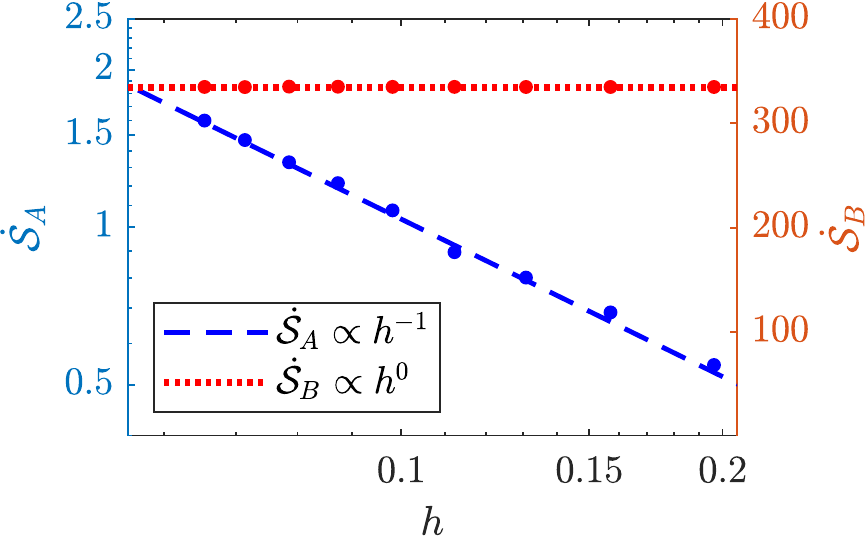}
    \caption{\textit{Scaling of $\dot{\mathcal{S}}_A$ and $\dot{\mathcal{S}}_B$ in lattice spacing $h$ for $\alpha=0.015$ ---} We observe that the two contributions to the entropy production rate in the ${\cal PT}$-symmetry-breaking phase of the two-species nonreciprocal Cahn-Hillard model scale differently with the lattice spacing, $h$. In particular, $\dot{\mathcal{S}}_A \propto h^{-1}$ and $\dot{\mathcal{S}}_B \propto h^{0}$. This implies that for small lattice spacing, the entropy production rate scales as $h^{-1}$ and thus diverges in the continuous limit, $h\rightarrow 0$.}
    \label{fig:epn}
\end{figure}

We run simulations varying the total number of lattice sites $M$ while keeping the system size $L$ constant to study the effect of the latter on the entropy production, which we compute via the It\^{o} integral Eq.\,\eqref{eq:conv1} as discussed above.
As expected, we find that the entropy production diverges in the limit $h \to 0$, implying that this is an ill-defined quantity in the continuum theory.
More precisely, we observe a linear scaling of the entropy production rate with the number of lattice sites, $\dot{\mathcal{S}}\propto M$ (equivalently, $\dot{\mathcal{S}} \propto h^{-1}$). These findings are shown in Fig.\,\ref{fig:epn}, where we represent separately the two contributions to the entropy production $\dot{\mathcal{S}}_A$ and $\dot{\mathcal{S}}_B$ as defined in Eqs.~\eqref{eq:Sa_SI} and \eqref{eq:SB_full_SI} below. 
 
 Interestingly, only the first of these appears to be responsible for the divergence, while the second contribution shows no dependence on $h$ and remains finite in the limit $h\to0$. This can be partly understood by interpreting the factor $h^{-1}$ in the definition of the lattice noise covariance, Eq.\,\eqref{eq:disc_noise_corr}, as a lattice-size dependent rescaling of the diffusion coefficient $D$ and recalling that the two contributions $\dot{\mathcal{S}}_A$ and $\dot{\mathcal{S}}_B$ scale differently with $D$ (see Sec.~\ref{s:lowD}), at least in the weak noise limit. 

\subsection{On Failure of Soft UV Regulators}
\label{sec:regUV}
The divergence of the entropy production rate \eqref{eq:ito_int} in the continuum limit $h \to 0$ should not come as a surprise since the former draws on equal-time, equal-position correlation functions of the fields and their derivatives. These are typically divergent when the power spectrum of the noise $\mathbf{\Lambda}$ is taken to have the usual flat form, Eq.\,\eqref{eq:disc_noise_corr}, which is of course unphysical \cite{Tauber2014}. While we have shown that recasting the dynamics on a regular lattice successfully regularizes this divergence, we may wonder whether a ``softer" regularization of the noise in the UV (e.g.\ one that ensures that its power spectrum decays exponentially at large wavenumbers and frequencies) is a sufficient measure to avoid divergences in the entropy production. Remarkably, this turns out not to be the case. 

To see this, consider again the continuum theory for the binary system of Cahn-Hilliard fields interacting through nonreciprocal couplings, as introduced in the main text. For simplicity, we focus on the homogeneous phase, where we can ignore the quartic term on the basis of fluctuations being small. We then write the simplified governing equations as
\begin{align}
    \dot{\phi}_1 = \nabla^2[- \phi_1 - \alpha \phi_2 + \gamma \nabla^2 \phi_1] + \Lambda_1 \nonumber \\
    \dot{\phi}_2 = \nabla^2[- \phi_2 + \alpha \phi_1 + \gamma \nabla^2 \phi_2] + \Lambda_2 \label{eq:lin_dyn_nrch}
\end{align}
with momentum-frequency covariance
\begin{equation}
    \langle \Lambda_k(q,\omega) \Lambda_{k'}(q',\omega')\rangle = 2D \delta_{kk'} q^2 \delta(q+q') \delta(\omega + \omega')  R^{(\omega)}(\omega^2) R^{(q)}(q^2) 
\end{equation}
where $R^{(q)}$ and $R^{(\omega)}$ denote UV regulators in momentum $q$ and frequency $\omega$. A common choice for these is $R^{(q)} = (1 + \xi^2q^2)^{-1}$ and $R^{(\omega)} = (1 + \tau^2\omega^2)^{-1}$, which introduced temporal correlations of the noise over a characteristic lengthscale $\xi$ and timescale $\tau$, which can be sent to zero to recover white noise. The entropy production for the regularized theory can be derived as done in Sec.~\ref{s:epr_fourier} to obtain, after taking the limit $h\to0$,
\begin{equation} \label{eq:regul_s}
    \dot{\mathcal{S}} \propto \int dq \int d\omega \left\langle \frac{\dot{\phi}_1(q,\omega) \phi_2(-q,-\omega)}{R^{(\omega)}(\omega^2) R^{(q)}(q^2)} - \frac{\dot{\phi}_2(q,\omega) \phi_1(-q,-\omega)}{R^{(\omega)}(\omega^2) R^{(q)}(q^2)} \right\rangle ~.
\end{equation}
Now, denoting $\langle ... \rangle_0$ expectations evaluated in the unregularized theory (where $R^{(q)}=R^{(\omega)}=1$) we have by linearity of the dynamics \eqref{eq:lin_dyn_nrch} that moments of order two in products of the fields and their derivatives satisfy 
\begin{equation}
    \left\langle q^n \phi_k(q,\omega) \phi_{k'}(-q,-\omega) \right\rangle = R^{(q)}(q^2) R^{(\omega)}(\omega^2)  \langle q^n \phi_k(q,\omega) \phi_{k'}(-q,-\omega) \rangle_0
\end{equation}
and their real-space representation can be made convergent by a suitable choice of the regulators $R^{(q,\omega)}$.
However, substituting into Eq.\,\eqref{eq:regul_s},
\begin{equation}
    \left\langle \frac{\dot{\phi}_1(q,\omega) \phi_2(-q,-\omega)}{R^{(\omega)}(\omega^2) R^{(q)}(q^2)} - \frac{\dot{\phi}_2(q,\omega) \phi_1(-q,-\omega)}{R^{(\omega)}(\omega^2) R^{(q)}(q^2)} \right\rangle = \left\langle \dot{\phi}_1(q,\omega) \phi_2(-q,-\omega) \right\rangle_0 - \left\langle \dot{\phi}_2(q,\omega) \phi_1(-q,-\omega) \right\rangle_0~,
\end{equation}
i.e.\ we find that the contribution from the regulators cancels out and we are left with the same (divergent) expression as in the unregularized theory. This hints at the fact that such divergences in the entropy production rate are in fact \emph{not} due to corresponding divergences in the correlations, which are amenable to regularization, but rather originate from the infinite dimensionality of the continuum field (as recently argued in \cite{Pruessner2022}) and can thus only be cured by imposing a finite dimensionality, e.g. through the introduction of a lattice.

\section{Full Details of Analytic Results in Weak-noise Regime}
\label{s:lowD}

Here, we return to the result in continuous space derived in full in Sec.\,\ref{sec:EPRcontspace}. For our model of nonreciprocally-coupled Cahn-Hilliard equations, we show that the entropy production rate Eq.\,(\ref{eq:final-epr}) can be written as the sum of two contributions which we interpret physically. Working in the weak noise limit, we derive analytic expressions for the leading order term for each contribution. This is done using the one-mode approximation presented in \cite{You2020}, which we also outline here for completeness.

\subsection{Expansion of Fields in Small Parameter $\sqrt{D}$}
The strength of the nonreciprocal coupling in the system considered is controlled by the parameter $\alpha$. Above a critical value $\alpha_c$, a pair of travelling wave solutions emerge. It is instructive to consider the re-formulated dynamics in the comoving frame of reference $(r', t') = (r-v(\alpha)t, t)$ where $v(\alpha)
$ is the phase velocity of the travelling wave. We define $\Phi_i(r', t') = \phi_i(r - v(\alpha)t, t)$ for $i=1, 2$ and expand each of our two fields $\Phi_1$ and $\Phi_2$ about the weak noise limit $D=0$, 
\begin{equation}\label{eq:Phi_exp_SI}
    \Phi_i(r', t') = \Phi_i^{(0)}(r') + \sqrt{D}\Phi_i^{(1)}(r', t') + D\Phi_i^{(2)}(r', t') + \mathcal{O}(D^{3/2}). 
\end{equation}
We choose to expand in powers of $\sqrt{D}$ as opposed to $D$ since the noise appears at order $\sqrt{D}$ in the Langevin equation. In the new frame of reference $(r', t')$, the deterministic contribution to the solution is stationary, $\dot{\Phi}_i^{(0)} = 0$. Note that, by construction, all functions $\Phi_i^{(j)}$ appearing on the right hand side of Eq.\,\eqref{eq:Phi_exp_SI} are independent of $D$. Below, we use the results 
\begin{equation}\label{eq:ssrel}
    \lim_{\tau\rightarrow\infty} \frac{1}{\tau}\int_0^t dt \langle \dot{\Phi}_i\rangle = \lim_{\tau\rightarrow\infty} \frac{1}{\tau}\int_0^t dt \langle \dot{\Phi}^{(j)}_i\rangle = 0
\end{equation}
to simplify the expressions for the entropy production.

Governing equations for the field perturbations $\Phi_i^{(j)}$ can be derived by substituting Eq.\,\eqref{eq:Phi_exp_SI} into the equations of motion for $\Phi_i(r', t')$ and collecting terms at each order in the expansion parameter $\sqrt{D}$. The terms of order $D^0$ recapitulate the deterministic nonreciprocal Cahn-Hilliard problem as studied in \cite{Saha2020,You2020}. In the comoving frame of reference, these deterministic solutions satisfy
\begin{align} \label{eq:det1}
    0&= \partial_{r'} \left[v(\alpha)\Phi_1^{(0)} + \partial_{r'}\mu^{(0)}_1+\alpha \Phi_2^{(0)} \right],\quad  \mu_1^{(0)} = \chi_1 \Phi_1^{(0)} + \frac{1}{3} \big(\Phi_1^{(0)}\big)^3 -\gamma_1 \partial_{r'}^2\Phi^{(0)}_1  \\
    0 &= \partial_{r'} \left[ v(\alpha)\Phi_2^{(0)} + \partial_{r'}\mu^{(0)}_2-\alpha\Phi_1^{(0)}\right],\quad  \mu_2^{(0)} = \chi_2 \Phi_2^{(0)}. \label{eq:det2}
\end{align}

At order $\sqrt{D}$, we derive the dynamics for the first order perturbations,
\begin{align} \label{eq:s34}
    \dot\Phi_1^{(1)} &= \partial_{r'} \left[v(\alpha)\Phi_1^{(1)} + \partial_{r'}\mu^{(1)}_1+\alpha \Phi_2^{(1)} + \sqrt{2}\Lambda_{1}\right],\quad  \mu_1^{(1)} = \chi_1 \Phi_1^{(1)} + \big(\Phi_1^{(0)}\big)^2\Phi_1^{(1)} -\gamma_1 \partial_{r'}^2\Phi^{(1)}_1  \\
    \dot{\Phi}_2^{(1)} &= \partial_{r'} \left[ v(\alpha)\Phi_2^{(1)} + \partial_{r'}\mu^{(1)}_2-\alpha\Phi_1^{(1)} + \sqrt{2}\Lambda_{2} \right],\quad  \mu_2^{(1)} = \chi_2 \Phi_2^{(1)} \label{eq:s35}
\end{align}
where the zero-mean white noise terms $ \Lambda_{i}(t)$ satisfying Eq.\,\eqref{eq:cont_noise_corr}. 

\subsection{Leading order contribution to $\dot{\mathcal{S}}_A$} 

Now, we turn to evaluating the leading-order terms in the contributions to the entropy production rate using the weak-noise expansion Eq.\,(\ref{eq:Phi_exp_SI}). In the main text, we derive the following expression for $\dot{\mathcal{S}}_A$ as a function of the two fields in the comoving frame:
\begin{equation}
    \dot{\mathcal{S}}_A = - \lim_{\tau\rightarrow\infty} \frac{\alpha}{D\tau}\int_0^\tau dt' \int_0^L dr'  \big\langle\Phi_2\dot\Phi_1 - \Phi_1\dot\Phi_2\big\rangle 
\end{equation}
We now substitute in the expansion Eq.\,\eqref{eq:Phi_exp_SI} and collect terms at leading order in $\sqrt{D}$ to write
\begin{equation}\begin{aligned} 
    \dot{\mathcal{S}}_A = -\lim_{\tau\rightarrow \infty}\frac{\alpha}{\tau L}\int_0^\tau dt'\int_0^Ldr'\:\bigg[&\frac{1}{D}\bigg(\Phi_2^{(0)}\dot{\Phi}_1^{(0)}-\Phi_1^{(0)}\dot{\Phi}_2^{(0)} \bigg)\\ 
    &+\frac{1}{\sqrt{D}}\bigg(\Phi_2^{(0)}\langle\dot{\Phi}_1^{(1)}\rangle-\Phi_1^{(0)}\langle\dot{\Phi}_2^{(1)}\rangle+\langle\Phi_2^{(1)}\rangle\dot{\Phi}_1^{(0)}-\langle\Phi_1^{(1)}\rangle\dot{\Phi}_2^{(0)} \bigg)\\ 
    &+ \langle\Phi_2^{(1)}\dot{\Phi}_1^{(1)}-\Phi_1^{(1)}\dot{\Phi}_2^{(1)}\rangle + \Phi_2^{(0)}\langle\dot{\Phi}_1^{(2)}\rangle-\Phi_1^{(0)}\langle\dot{\Phi}_2^{(2)}\rangle+\langle\Phi_2^{(2)}\rangle\dot{\Phi}_1^{(0)}-\langle\Phi_1^{(2)}\rangle\dot{\Phi}_2^{(0)}  
    +\mathcal{O}(D^{1/2})\bigg]~.
\end{aligned}\end{equation} 
This simplifies significantly when we recall both the relation Eq.\,(\ref{eq:ssrel}) and that $\dot\Phi_i^{(0)}=0$. The leading order contribution to $\dot{\mathcal{S}}_A$ thus takes the form 
\begin{equation}\label{eq:Sa_SI} 
    \dot{\mathcal{S}}_A = -\lim_{\tau\rightarrow \infty}\frac{\alpha}{\tau L}\int_0^\tau dt'\int_0^Ldr'\: \left\langle\Phi_2^{(1)}\dot{\Phi}_1^{(1)}-\Phi_1^{(1)}\dot{\Phi}_2^{(1)}\right\rangle + \mathcal{O}(D^{1/2}). 
\end{equation} 
A closed analytic form for the right hand side of \eqref{eq:Sa_SI} requires solving Eqs.\,\eqref{eq:s34} and \eqref{eq:s35}, but in general these equations do not admit analytic solutions.

\subsection{Leading order contribution to $\dot{\mathcal{S}}_B$} 
The remaining contribution $\dot{\mathcal{S}}_B$ was defined in the main text as
\begin{equation}
    \dot{\mathcal{S}}_B = \lim_{\tau\rightarrow\infty} \frac{\alpha v(\alpha)}{D\tau }\int_0^\tau dt' \int_0^L dr' \big\langle\Phi_2\partial_{r'}\Phi_1 - \Phi_1\partial_{r'}\Phi_2\big\rangle.
\end{equation}
Again, we substitute the expansion in Eq.\,(\ref{eq:Phi_exp_SI}) into this expression to write the leading-order
\begin{equation}\label{eq:SB_full_SI} 
    \dot{\mathcal{S}}_B = \lim_{\tau\rightarrow\infty}\frac{\alpha v(\alpha)}{D\tau L}\int_0^\tau dt'\int_0^Ldr'\:  \bigg(\Phi_2^{(0)}\partial_{r'}\Phi_1^{(0)} - \Phi_1^{(0)}\partial_{r'}\Phi_2^{(0)}\bigg) + \mathcal{O}(D^{-1/2}). 
\end{equation} 
We recognize this integral as the global polar order parameter, evaluated for the deterministic solutions \cite{Saha2020}
\begin{equation}\label{eq:J0_SI} 
    \mathcal{J}^{(0)} = \frac{1}{\tau L}\int_0^\tau dt'\int_0^Ldr'\: \bigg( \Phi_2^{(0)}\partial_{r'}\Phi_1^{(0)} - \Phi_1^{(0)}\partial_{r'}\Phi_2^{(0)}\bigg). 
\end{equation} 
The current $\mathcal{J}^{(0)}$, for which we rederive an explicit form in the following section though a one-mode approximation, is non-zero when $\mathcal{PT}$-symmetry is broken in the deterministic dynamics. This allows us to write the second contribution to the entropy production rate $\dot{\mathcal{S}}_B$ compactly as $\dot{\mathcal{S}}_B = {\alpha v(\alpha)\mathcal{J}^{(0)}}/{D}$.

\subsection{One-mode Approximation and Closed-form Expression for $\dot{\mathcal{S}}_B$} 

Here, we briefly recap the results of \cite{You2020}, where the deterministic problem Eqs.\,\eqref{eq:det1} and \eqref{eq:det2} was studied for $L=2\pi$ and $|\chi_1/\gamma_1| \gtrsim (2\pi/L)^2$. With this choice of parameters and looking at the dispersion relation, one can argue that only the lowest non-zero wavenumber mode $q=2\pi/L$ is linearly unstable. This is consistent with the observation that the solutions $\Phi_i^{(0)}$ (measured through numerical simulations) were largely dominated by the lowest (finite) wave-number Fourier mode. In this case, we can thus proceed to a \textit{one-mode approximation}, 
\begin{equation} \label{eq:onemodapprox}
     \Phi_j^{(0)}(r) = 
    \sum_{n=-\infty}^\infty \bar\zeta_j(n) e^{i n r} 
    \approx \bar\zeta_j(-1) e^{-i r} + \bar\zeta_j(+1) e^{ i r} = \zeta_j \cos( r-\theta_j),
\end{equation} 
where we used $\bar\zeta_j(+1) = \bar\zeta^*_j(-1) = \zeta_j e^{-i \theta_j}$ with $\zeta_i \in \mathbb{R}$ for a real function $\Phi_j^{(0)}$;
this can be shown to agree very accurately with the full solutions for the stationary distributions in the present case but we do not expect it to hold more generally \cite{FrohoffHulsmann2021}. 

Note that the homogeneous mode disappears because we have set the average value of each field to zero. Substituting Eq.\,\eqref{eq:onemodapprox} into the deterministic equations of motion, four governing equations can be derived for the coefficients $\zeta_1, \zeta_2, \theta_1$ and $\theta_2$ \cite{You2020}. The steady-state solutions to these dynamic equations can then be found and three sets of solutions can be identified, corresponding to a trivial fixed point where the fields each are equivalently zero, the static phase separated state and the phase separated state that supports a travelling solution. 

\begin{figure}
    \centering
    \includegraphics{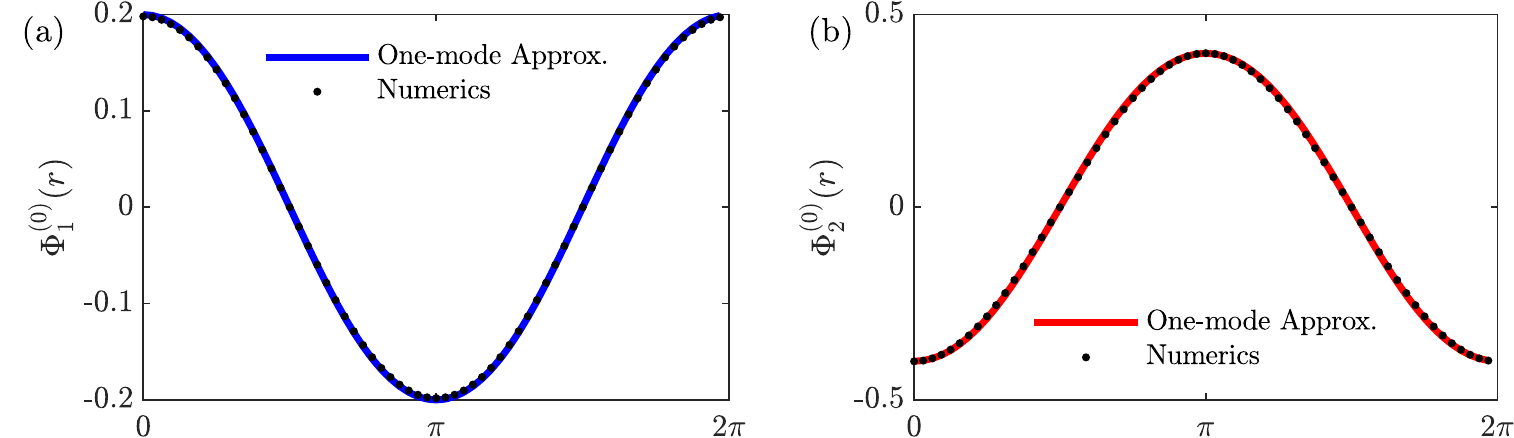}
    \caption{\textit{One-mode approximation for deterministic solutions --- }we compare the numerical solutions to the deterministic equations Eqs.\,\eqref{eq:det1} and \eqref{eq:det2} to the results of \cite{You2020} where a one-mode approximation is used for an analytic form for the deterministic solutions for $L=2\pi$. We plot the results for the two fields, (a) $\Phi_1^{(0)}$ and (b) $\Phi_2^{(0)}$.}
    \label{fig:1mode}
\end{figure}

The sets of non-trivial solutions are what we use to find an approximate analytic form for $\Phi_j^{(0)}$, being careful to perform the change of reference frame $(r, t)\rightarrow (r', t')$ from the results of \cite{You2020}. We eventually obtain
\begin{align} 
    \Phi_1^{(0)}(r') &= {2\sqrt{-\chi_1-\gamma_1-\chi_2}}\cos( r') \nonumber \\
    \Phi_2^{(0)}(r')&=\frac{2\sqrt{(\alpha+\kappa)(-\chi_1-\gamma_1-\chi_2)}}{\sqrt{\alpha-\kappa}}\cos\bigg(r'-\arccos\bigg(-\sqrt{{\chi_2^2}/(\alpha^2-\kappa^2})\bigg)\Bigg)  \label{eq:an_onmod}
\end{align} 
which define a one-parameter class of solutions by the translational symmetry of the problem through the transformation $r'\rightarrow r' + \Delta r'$. Note that, for the travelling wave solution to exist, we require $\chi_1+ \gamma_1<-\chi_2$ and $\alpha^2\geq \alpha_c^2 = \kappa^2 + \chi_2^2$. In Fig. \ref{fig:1mode}, we compare these approximate analytic solutions to the deterministic solutions evaluated via numerical simulations, observing good agreement as reported in \cite{You2020}.

Substituting Eq.\,\eqref{eq:an_onmod} into Eq.\,(\ref{eq:J0_SI}), we obtain a closed form expression for the deterministic global polar order parameter $\mathcal{J}^{(0)}$,
\begin{equation} 
    \mathcal{J}^{(0)} = \frac{8\pi v(\alpha)(-\chi_1-\gamma_1-\chi_2)}{(\alpha-\kappa)}\quad \text{where}\quad v(\alpha) = \sqrt{\alpha^2-\kappa^2-\chi_2^2}, 
\end{equation} 
and thus, via Eq.\,\eqref{eq:SB_full_SI}, an expression for the leading order contribution to $\dot{\mathcal{S}}_B$,
\begin{equation}\label{eq:SB_full_SI} 
	\dot{\mathcal{S}}_B = \frac{8\pi\alpha v^2(\alpha)(-\chi_1-\gamma_1-\chi_2)}{D(\alpha-\kappa)}.
\end{equation} 
Note that the conditions for the existence of the traveling wave solution imply $\dot{\mathcal{S}}_B\geq 0$ in agreement with the second law of thermodynamics. For $\alpha \gg \alpha_c$, we observe the scaling relation $\dot{\mathcal{S}}_B = {v^2(\alpha)}/{D_{\rm eff}}$ with $D_{\rm eff} = D(\kappa-\alpha)/(8\pi\alpha(\chi_1+\gamma_1+\chi_2))$. In this limit, Eq.\,\eqref{eq:SB_full_SI} for the contribution to the entropy production stemming from the macroscopic dynamic phase thus bears a striking resemblance to the entropy production of an active particle on a ring with bare diffusivity $D_{\rm eff}>0$ and self-propulsion speed $v(\alpha)$ \cite{Cocconi2020, Seifert2012}.  


%